**Title**

# Large Language Model–Assisted Cleaning of Report-Derived Labels in a Large-Scale Chest CT Dataset

**Authors:**

- Yosuke Yamagishi, MD, MSc
  Division of Radiology and Biomedical Engineering, Graduate School of Medicine, The University of Tokyo, Tokyo, Japan
- Atsushi Takamatsu, MD, PhD
  Department of Computational Diagnostic Radiology and Preventive Medicine, The University of Tokyo Hospital, Tokyo, Japan
  Department of Radiology, Kanazawa University Hospital, Ishikawa, Japan
- Mototsugu Sato, MEng
  Faculty of Medicine, The University of Tokyo, Tokyo, Japan
- Tomohiro Kikuchi, MD, PhD, MPH
  Department of Computational Diagnostic Radiology and Preventive Medicine, The University of Tokyo Hospital, Tokyo, Japan
  Department of Radiology, School of Medicine, Jichi Medical University, Tochigi, Japan
- Shouhei Hanaoka, MD, PhD
  Division of Radiology and Biomedical Engineering, Graduate School of Medicine, The University of Tokyo, Tokyo, Japan
  Department of Radiology, The University of Tokyo Hospital, Tokyo, Japan
- Takeharu Yoshikawa, MD, PhD
  Department of Computational Diagnostic Radiology and Preventive Medicine, The University of Tokyo Hospital, Tokyo, Japan
- Osamu Abe, MD, PhD
  Division of Radiology and Biomedical Engineering, Graduate School of Medicine, The University of Tokyo, Tokyo, Japan
  Department of Radiology, The University of Tokyo Hospital, Tokyo, Japan

**Corresponding Author:**

Yosuke Yamagishi, MD, MSc

Email: yamagishi-yosuke0115@g.ecc.u-tokyo.ac.jp

## Abstract

**Purpose:** To evaluate whether large language model (LLM)–assisted label cleaning can identify label–report discordance in CT-RATE, a large-scale public chest CT dataset.

**Materials and Methods:** After report-level deduplication, 24,446 unique radiology reports were identified. Twelve reports were excluded from the primary GPT-5.4 analysis because of Microsoft Azure AI Foundry content-safety filtering, leaving 24,434 reports and 439,812 label instances across 18 abnormality categories. GPT-5.4-derived binary labels were generated from report text using structured JSON output and compared with existing CT-RATE labels. Discordant instances were adjudicated by radiologists. In addition, 100 randomly sampled reports were manually annotated to compare CT-RATE labels, individual LLM-derived labels, and multi-LLM majority-vote labels against radiologist-annotated reference labels.

**Results:** Overall agreement between GPT-5.4-derived and CT-RATE labels was 96.4%, with Cohen's κ of 0.884. Lymphadenopathy showed the lowest agreement and κ. In discordance review, radiologist adjudication supported GPT-5.4-derived labels in 72 of 97 (74.2%) general discordant instances and 91 of 99 (91.9%) targeted lymphadenopathy discordant instances. Against radiologist-annotated reference labels, multi-LLM majority-vote labels achieved the highest label-macro-averaged F1 score and Cohen's κ.

**Conclusion:** LLM-assisted label cleaning identified clinically meaningful label–report discordance in CT-RATE and may support scalable quality improvement of public imaging datasets. The cleaned dataset will be made publicly available to support future research.

## Introduction

Public medical imaging datasets are increasingly used to benchmark artificial intelligence models and to train new imaging algorithms (1,2). Because benchmark performance is often interpreted as evidence of model capability, the quality of structured labels in these datasets is critically important. Even in carefully constructed public datasets, label errors and inconsistencies can occur because of reporting style, negation, uncertainty, ambiguous label definitions, and automated extraction procedures (3,4).

CT-RATE is a valuable large-scale chest CT dataset that includes volumetric CT images, radiology reports, and multi-abnormality labels (5). Its labels have enabled the development and evaluation of three-dimensional chest CT models. However, as with other large-scale report-derived datasets, some structured labels may be discordant with the evidence contained in the corresponding radiology reports. Because public datasets are repeatedly reused for benchmarking, even modest label errors can propagate into model evaluation and comparison. Scalable label cleaning is therefore an important component of dataset maintenance.

Large language models can perform structured information extraction from clinical text and may provide a practical tool for identifying potential label–report discordance (6,7). Compared with rule-based or task-specific extraction systems, LLM-based approaches may reduce the need for task-specific feature engineering or model training and can be configured to return structured outputs across multiple categories (8,9). However, the extent to which LLM-derived labels agree with existing dataset labels, and whether discordant cases reflect errors in existing labels or limitations of LLM extraction, remains uncertain.

The purpose of this study was to evaluate whether an LLM-assisted framework could identify and characterize discordance between structured abnormality labels and radiology report text in a public chest CT dataset, with the additional aim of releasing a cleaner set of report-aligned labels.

## Materials and Methods

The overall workflow of the LLM-assisted label cleaning framework is summarized in Figure 1.

### Dataset

This retrospective study used CT-RATE, a publicly available chest CT dataset containing CT volumes, radiology reports, and structured abnormality labels (5). CT-RATE was selected because it is, to our knowledge, the largest publicly available chest CT dataset paired with radiology reports and is released under a CC BY-NC-SA license, permitting release of derived labels for noncommercial research. Because CT-RATE includes multiple reconstructed CT volumes from the same CT examination, we performed report-level deduplication. Among 24,446 unique reports, 22,976 were from the training set and 1,470 from the validation set. The analysis included the 18 predefined CT-RATE abnormality categories. Reports blocked by Microsoft Azure AI Foundry content-safety filtering were excluded from the corresponding analyses.

### LLM-Assisted Label Cleaning

Each radiology report was processed using GPT-5.4 through Microsoft Azure AI Foundry (10), selected because Microsoft states that customer prompts and completions are not used to train or improve Microsoft, provider, or third-party models. The model assigned binary labels for the 18 predefined CT-RATE abnormality categories using only the findings and impression sections. Findings clearly described as present, observed, or detected were labeled present, whereas explicitly negated or unmentioned findings were labeled absent. Uncertain findings, such as possible or suspected abnormalities, were treated as present in the primary extraction. The prompt instructed the model not to infer findings from clinical context, epidemiology, or general medical knowledge. Outputs were required in structured JSON format, and invalid or incomplete outputs were reprocessed. LLM-derived labels were then compared with the original CT-RATE labels.

**Secondary LLMs and Multi-LLM Majority Voting**

To assess whether multi-model labeling improved reliability, the same reports were additionally processed using GPT-5.4 mini and DeepSeek V3.2 (11) with the same prompt and output format. These models were selected to complement GPT-5.4, the primary state-of-the-art model available through Azure AI Foundry, with its smaller variant and a state-of-the-art open-source model. GPT-5.4 was predefined as the primary model for discordance analysis and adjudication sampling. In the random report-level reference set, GPT-5.4, GPT-5.4 mini, DeepSeek V3.2, and majority-vote labels were compared with radiologist-confirmed annotations. Majority-vote labels were assigned when at least two of the three LLMs selected the same class. Original CT-RATE labels, individual LLM labels, and majority-vote labels were evaluated against manual annotations.

**Label Comparison**

For each abnormality category, GPT-5.4-derived labels were compared with existing CT-RATE labels. Agreement and discordance were calculated at the label-instance level. Discordant cases were categorized as existing-label positive/GPT-5.4 negative or existing-label negative/GPT-5.4 positive.

**Radiologist Adjudication**

Manual validation consisted of three complementary review sets. First, 100 randomly sampled reports were manually annotated for all 18 predefined abnormality categories, yielding 1,800 radiologist-confirmed label instances for comparing original CT-RATE labels, individual LLM-derived labels, and multi-LLM majority-vote labels against manual annotations. Second, 100 general discordant label instances were sampled from CT-RATE/GPT-5.4 disagreements, including 50 CT-RATE positive/GPT-5.4 negative and 50 CT-RATE negative/GPT-5.4 positive instances. Labels selected for targeted review were excluded from this set. Third, using the Landis and Koch scale as a conventional reference, labels not reaching “almost perfect agreement” (Cohen’s $\kappa \leq 0.80$) were selected for targeted discordance review (12). For each selected label, up to 50 discordant instances in each direction were sampled.
Initial annotation and adjudication were performed by a radiology fellow (postgraduate year 6). In the report-level reference set, all 18 categories were annotated from the report text. In the discordance review sets, the reviewer determined whether the CT-RATE label, GPT-5.4-derived label, both, or neither was consistent with the report. All annotations and adjudications were reviewed and finalized by a board-certified radiologist, with disagreements or uncertain cases resolved by consensus.

**Statistical Analysis**

Agreement and discordance rates between GPT-5.4-derived labels and CT-RATE labels were summarized by abnormality category. In the random report-level reference set, the original CT-RATE labels, individual LLM-derived labels, and multi-LLM majority-vote labels were compared with radiologist-confirmed manual annotations. Performance was summarized using sensitivity, specificity, positive predictive value, negative predictive value, accuracy, F1 score, and Cohen's κ. Metrics were calculated for each abnormality category and summarized using label-macro averaging. Micro-averaged metrics were calculated by pooling true-positive, false-positive, true-negative, and false-negative label instances across all 18 categories.

Radiologist adjudication results for discordant cases were reported as proportions with 95% confidence intervals using the Wilson method. For discordance adjudication, the proportion of instances in which the GPT-5.4-derived label was supported was tested against a null proportion of 0.50 using an exact binomial test.

## Results

**Study Dataset**

After report-level deduplication, 24,446 unique radiology reports were identified. Twelve reports were excluded because they could not be processed by GPT-5.4 owing to Microsoft Azure AI Foundry content-safety filtering. The primary GPT-5.4 comparison therefore included 24,434 reports and 439,812 label instances across 18 predefined abnormality categories.

**Agreement Between CT-RATE Labels and GPT-5.4-Derived Labels**

Overall agreement between the original CT-RATE labels and GPT-5.4-derived labels was 96.4%, with a Cohen's κ of 0.884. Category-level agreement ranged from 79.4% for lymphadenopathy to 99.6% for hiatal hernia. Lymphadenopathy showed the lowest Cohen's κ among the 18 categories, with a κ of 0.309. Category-level agreement metrics are summarized in Table 1, and per-label confusion matrices are shown in Supplementary Figure 1.

**Discordance Review**

In the general discordance review set excluding lymphadenopathy, radiologist adjudication supported the GPT-5.4-derived label in 72 of 97 adjudicable instances excluding three instances in which both labels were supported (74.2%; 95% CI, 64.7–81.9), supported the CT-RATE label in 25 instances, and supported both labels in three instances. Across abnormality categories, adjudication more frequently supported the GPT-5.4-derived labels than the original CT-RATE labels for most categories (Figure 2). Lymphadenopathy was the only category with Cohen's κ ≤ 0.80 and was therefore selected for targeted discordance review. In this targeted review, adjudication supported the GPT-5.4-derived label in 91 of 99 instances (91.9%; 95% CI, 84.9–95.8) and the CT-RATE label in 8 instances, with one instance supporting both labels. Radiologist adjudication results are summarized in Supplementary Table 1.

### Validation Against Radiologist-Annotated Reference Labels

In the radiologist-annotated reference set, 100 randomly sampled reports were manually annotated for all 18 abnormality categories, yielding 1,800 reference label instances. Compared with these reference labels, the original CT-RATE labels achieved a label-macro-averaged F1 score of 93.8% and Cohen's κ of 0.923, whereas GPT-5.4-derived labels achieved an F1 score of 95.1% and Cohen's κ of 0.940. Multi-LLM majority-vote labels achieved the highest label-macro-averaged F1 score of 96.0% and Cohen's κ of 0.951. Notably, majority voting resulted in Cohen's κ values exceeding 0.80 for all 18 abnormality categories, corresponding to almost perfect agreement, while GPT-5.4-derived labels achieved this threshold for 17 of 18 categories. Per-label F1 scores and Cohen's κ values are shown in Figure 3, and label-macro-averaged performance metrics are summarized in Table 2.

## Discussion

This study evaluated an LLM-assisted framework for cleaning report-derived labels in CT-RATE, a large-scale public chest CT dataset. GPT-5.4-derived labels showed high aggregate agreement with existing CT-RATE labels, but category-level agreement varied substantially, ranging from almost perfect agreement for 17 of 18 categories to fair agreement for lymphadenopathy (κ = 0.309). Radiologist adjudication of discordant cases supported the GPT-5.4-derived labels in most reviewed instances, both in the general discordance set and in the targeted lymphadenopathy review. In the radiologist-annotated reference set, GPT-5.4-derived labels slightly exceeded the original CT-RATE labels in aggregate performance, and multi-LLM majority-vote labels achieved the highest overall performance. These findings suggest that LLM-assisted label cleaning can identify label–report discordance that may not be apparent from aggregate agreement metrics alone and that multi-model labeling may further improve the reliability of refined report-derived labels.

The results support the broader concept that public medical imaging datasets require ongoing label quality assessment. Public datasets are repeatedly used for benchmarking, method comparison, and model development (4,13,14). Therefore, even modest label inconsistencies can affect performance estimates, model ranking, and interpretation of model capability.

The most notable category-specific finding was lymphadenopathy, which showed substantially lower agreement than all other abnormality categories. This finding is clinically important because lymphadenopathy is central to lung cancer staging and malignancy assessment. Thoracic lymphadenopathy is also encountered in a broad range of infectious, inflammatory, neoplastic, and systemic diseases (15,16). Poor reliability of this label may therefore have disproportionate effects on both benchmark interpretation and downstream model development.

Qualitative review suggested that lymphadenopathy discordance often reflected ambiguous mapping between report language and the structured label. Some CT-RATE positive/GPT-5.4 negative cases appeared to contain only nonspecific references to small or millimetric lymph nodes, rather than pathologic lymph node enlargement. Other cases involved abbreviations such as “LAP,” negated expressions, or size-based descriptions that required interpretation of pathologic thresholds and radiologic context. This flexibility is consistent with prior studies showing that LLMs can perform zero- or few-shot structured information extraction from clinical text and transform free-text radiology reports into structured outputs with limited task-specific training (17,18). More explicit criteria distinguishing lymph node mention from clinically significant lymphadenopathy may further improve label consistency in future dataset updates.

More broadly, category-level discordance analysis may help identify labels that require refined definitions or targeted expert adjudication. The value of LLM-assisted label cleaning may therefore extend beyond correcting individual labels: it can also function as a dataset quality assessment tool that highlights label categories with ambiguous definitions, inconsistent extraction behavior, or high clinical importance. This approach may be applicable to other report-derived imaging datasets, provided that modality-specific label definitions and clinical thresholds are carefully validated. However, LLM-generated labels may themselves introduce bias when used as reference labels (19), underscoring the need for human verification in smaller validation subsets and for carefully curated evaluation labels, as emphasized in prior public imaging challenge datasets (20).

This study has limitations. First, LLM-derived labels and radiologist adjudication were based on report text and do not represent image-level ground truth. Second, the discordance review sets were enriched for disagreement and therefore should not be interpreted as direct estimates of overall dataset accuracy. Third, results may depend on the model version, prompt design, and label definitions. Future work should assess the effect of refined labels on model development and evaluate whether targeted cleaning of clinically important labels such as lymphadenopathy improves benchmark reliability.

In conclusion, LLM-assisted label cleaning identified label–report discordance in CT-RATE and helped prioritize clinically important categories, particularly lymphadenopathy, for expert review. This approach may provide a scalable framework for maintaining and improving the reliability of public imaging datasets used for AI benchmarking and model development.

## Data Availability

Data used in this study were obtained from the publicly available CT-RATE dataset (https://huggingface.co/datasets/ibrahimhamamci/CT-RATE). The refined labels generated in this study and the code used for cohort construction, large language model–based extraction, and downstream analyses will be made publicly available.

## Acknowledgements

The authors used ChatGPT (OpenAI) to assist with language editing and refinement of the manuscript. The authors reviewed and edited the output as needed and take full responsibility for the content of the final manuscript.

In this research work, we used the UTokyo Azure (https://utelecon.adm.u-tokyo.ac.jp/en/research_computing/utokyo_azure/) for the LLM-based extraction pipeline.

## Figures

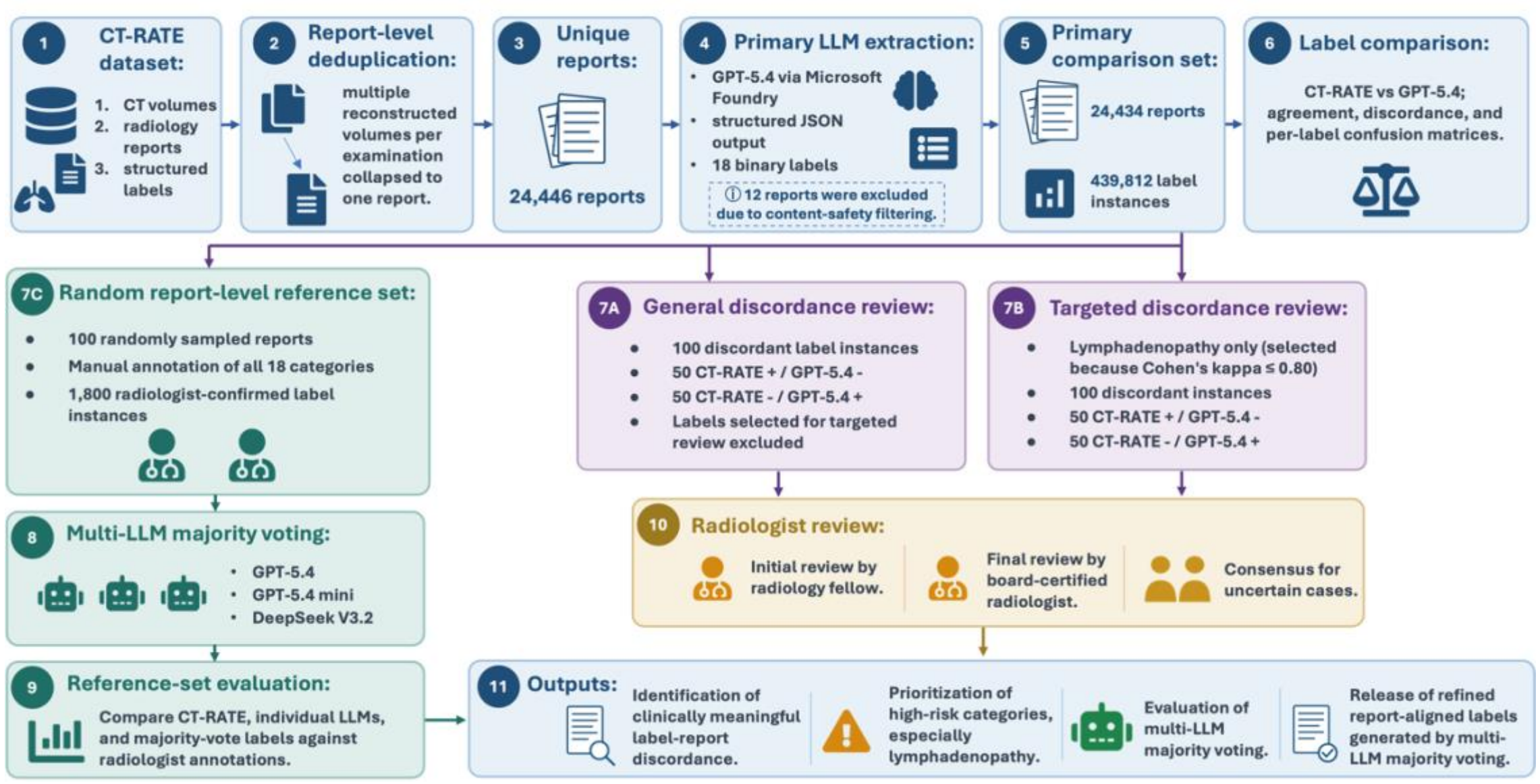

**Figure 1.** Workflow of the LLM-assisted label-cleaning and validation framework. After report-level deduplication of CT-RATE, unique radiology reports were processed with GPT-5.4 to generate structured binary labels for 18 abnormality categories. GPT-5.4-derived labels were compared with original CT-RATE labels to quantify agreement and identify discordant instances. Validation included a random report-level reference set, general discordance review, targeted lymphadenopathy discordance review, and multi-LLM majority-vote evaluation. Refined report-aligned labels generated by applying multi-LLM majority voting to all eligible reports will be released for public research use.

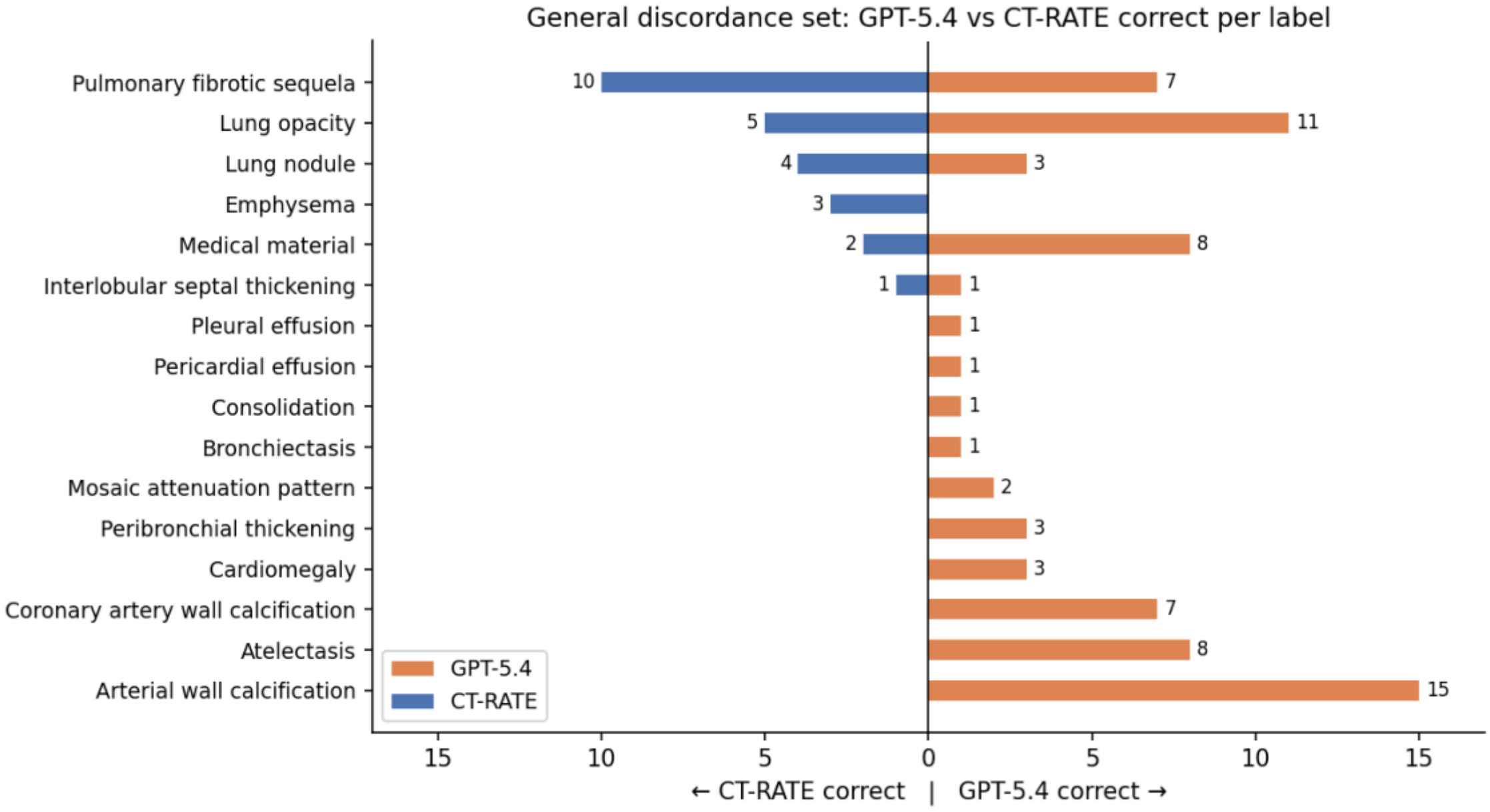

**Figure 2.** Radiologist adjudication results in the general discordance review set excluding lymphadenopathy. The figure shows the number of discordant label instances for which radiologist adjudication supported the GPT-5.4-derived label or the original CT-RATE label, stratified by abnormality category. Bar length indicates the number of discordant instances adjudicated in favor of each source. Rows are sorted by the number of instances adjudicated in favor of CT-RATE in descending order; categories with no CT-RATE-adjudicated instances are sorted by GPT-5.4-adjudicated count in ascending order. Hiatal hernia was excluded as no discordant instances were sampled due to its low overall discordance rate.

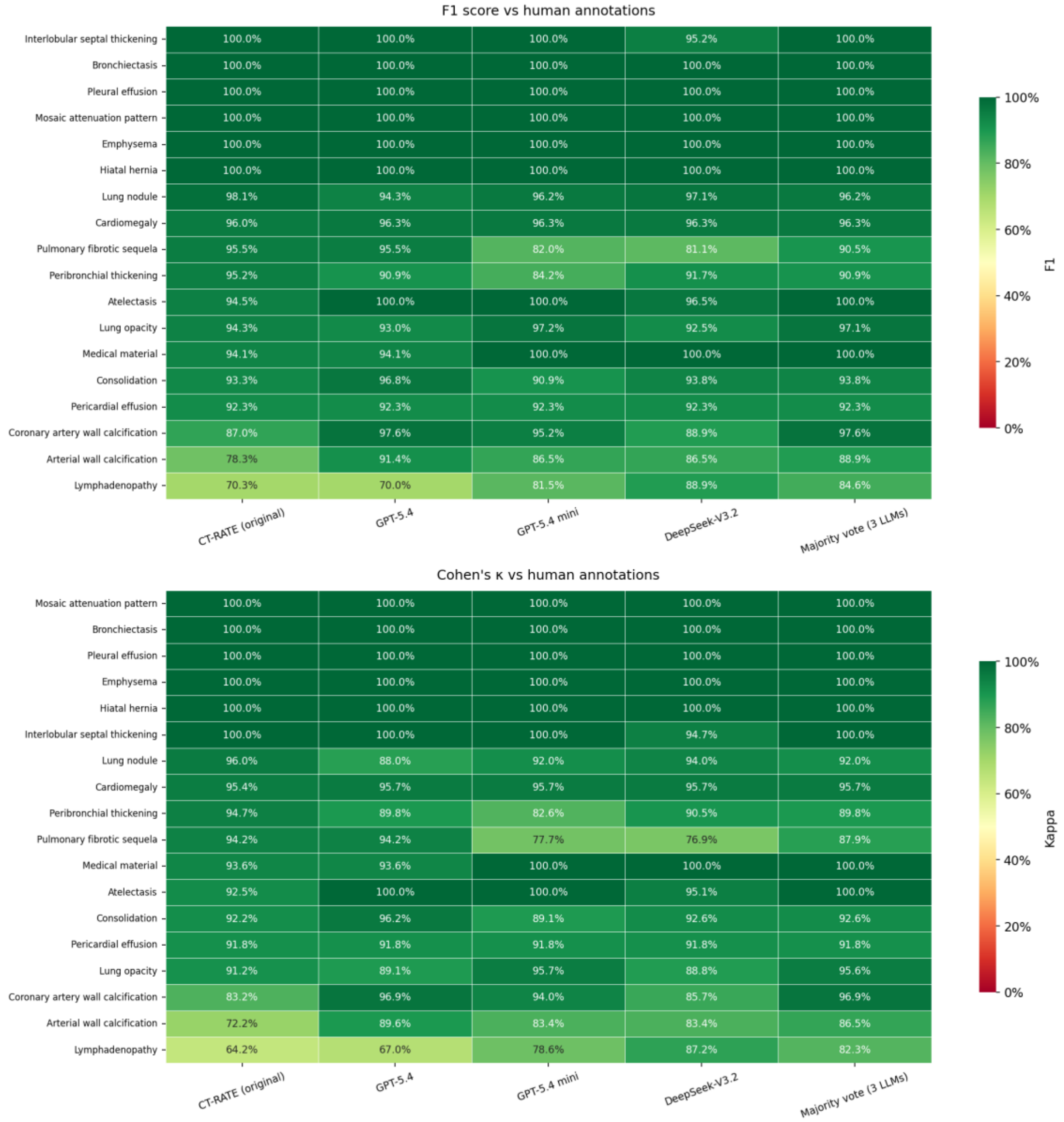

**Figure 3.** Per-label F1 scores and Cohen's κ against radiologist-annotated reference labels. Heatmap of per-label values for the original CT-RATE labels, individual LLM-derived labels, and multi-LLM majority-vote labels against radiologist-annotated reference labels. Rows indicate abnormality categories sorted in descending order of CT-RATE performance, and columns indicate label sources. Values in each cell indicate the F1 score and the Cohen's κ for the corresponding abnormality category and label source.

## Tables

**Table 1.** Per-label agreement between CT-RATE labels and GPT-5.4-derived labels. Values were calculated at the report level for each abnormality category. Overall agreement, positive agreement, negative agreement, discordance rate, and Cohen’s κ were calculated between the original CT-RATE labels and GPT-5.4-derived labels. N = 24,434 reports for each abnormality category. Rows are sorted by Cohen's κ in descending order.

| Label | Overall agreement (95% CI) | Positive agreement | Negative agreement | Discordance rate | Cohen's κ |
|---|---|---|---|---|---|
| Hiatal hernia | 99.6% (99.6–99.7) | 98.8% | 99.8% | 0.4% | 0.986 |
| Emphysema | 99.0% (98.9–99.2) | 97.7% | 99.4% | 1.0% | 0.971 |
| Pleural effusion | 99.2% (99.1–99.3) | 97.0% | 99.6% | 0.8% | 0.965 |
| Cardiomegaly | 99.2% (99.1–99.3) | 96.5% | 99.6% | 0.8% | 0.961 |
| Bronchiectasis | 99.2% (99.1–99.3) | 96.3% | 99.6% | 0.8% | 0.958 |
| Consolidation | 98.7% (98.5–98.8) | 96.5% | 99.2% | 1.3% | 0.956 |
| Pericardial effusion | 99.4% (99.3–99.5) | 95.7% | 99.7% | 0.6% | 0.954 |
| Mosaic attenuation pattern | 99.3% (99.1–99.4) | 95.5% | 99.6% | 0.7% | 0.951 |
| Atelectasis | 96.6% (96.3–96.8) | 93.9% | 97.6% | 3.4% | 0.915 |
| Lung nodule | 95.6% (95.3–95.9) | 95.5% | 95.7% | 4.4% | 0.912 |
| Interlobular septal thickening | 98.5% (98.4–98.7) | 91.5% | 99.2% | 1.5% | 0.907 |
| Medical material | 97.8% (97.6–97.9) | 91.2% | 98.7% | 2.2% | 0.899 |
| Lung opacity | 94.4% (94.1–94.7) | 92.7% | 95.5% | 5.6% | 0.881 |
| Coronary artery wall calcification | 95.2% (94.9–95.4) | 89.8% | 96.8% | 4.8% | 0.867 |
| Pulmonary fibrotic sequela | 94.6% (94.3–94.8) | 90.0% | 96.3% | 5.4% | 0.863 |
| Peribronchial thickening | 96.7% (96.5–96.9) | 86.0% | 98.1% | 3.3% | 0.842 |
| Arterial wall calcification | 92.7% (92.4–93.1) | 85.9% | 95.1% | 7.3% | 0.811 |
| Lymphadenopathy | 79.4% (78.9–79.9) | 38.4% | 87.7% | 20.6% | 0.309 |

**Table 2.** Performance against radiologist-annotated reference labels. Performance metrics for the original CT-RATE labels, individual LLM-derived labels, and multi-LLM majority-vote labels against radiologist-annotated reference labels in 100 randomly sampled reports. The reference set contained 1,800 label instances across 18 abnormality categories. PPV = positive predictive value; NPV = negative predictive value; LLM = large language model.

| Source | Sensitivity | Specificity | PPV | NPV | Accuracy | F1 | Cohen's κ |
|---|---|---|---|---|---|---|---|
| CT-RATE (original) | 96.3% (93.8–98.7) | 97.7% (95.6–99.8) | 92.9% (86.1–99.7) | 99.2% (98.6–99.8) | 97.6% (95.9–99.2) | 93.8% (89.8–97.8) | 0.923 (0.874–0.972) |
| GPT-5.4 | 94.4% (88.8–100.0) | 98.8% (97.7–99.9) | 96.8% (94.9–98.7) | 98.9% (98.0–99.8) | 98.4% (97.4–99.4) | 95.1% (91.6–98.6) | 0.940 (0.900–0.980) |
| GPT-5.4 mini | 94.8% (90.3–99.3) | 98.5% (97.6–99.5) | 94.9% (91.4–98.4) | 99.0% (98.2–99.9) | 98.2% (97.1–99.3) | 94.6% (91.2–97.9) | 0.934 (0.894–0.973) |
| DeepSeek-V3.2 | 95.5% (91.7–99.4) | 98.6% (97.8–99.4) | 94.0% (90.6–97.3) | 98.7% (97.6–99.8) | 97.9% (96.9–99.0) | 94.5% (91.7–97.2) | 0.931 (0.898–0.965) |
| Majority vote (3 LLMs) | 96.4% (93.7–99.2) | 98.8% (98.0–99.6) | 95.7% (93.1–98.3) | 99.3% (98.8–99.8) | 98.6% (97.8–99.4) | 96.0% (93.6–98.4) | 0.951 (0.923–0.979) |

## Supplementary Materials

**Supplementary Table 1**. Radiologist adjudication of discordant label instances. The table summarizes adjudication results for the general discordance review set, which excluded lymphadenopathy, and the focus discordance review set for lymphadenopathy. CT−/GPT+ indicates CT-RATE negative/GPT-5.4 positive discordance, and CT+/GPT− indicates CT-RATE positive/GPT-5.4 negative discordance. GPT correct, CT-RATE correct, both, and neither indicate the adjudication category assigned by radiologists. Percentages and 95% confidence intervals indicate the proportion of instances in which the GPT-5.4-derived label was adjudicated as correct. P values were calculated using an exact binomial test against a null proportion of 0.50.

| set | group | GPT correct | CT-RATE correct | both | neither | GPT correct (95%CI) | P value |
|---|---|---|---|---|---|---|---|
| General discordance | Overall (n=100) | 72 | 25 | 3 | 0 | 74.2% (64.7–81.9) | <0.001 |
| | CT−/GPT+ (n=50) | 37 | 13 | 0 | 0 | 74.0% (60.4–84.1) | <0.001 |
| | CT+/GPT− (n=50) | 35 | 12 | 3 | 0 | 74.5% (60.5–84.7) | 0.001 |
| Focus discordance (Lymphadenopathy) | Overall (n=100) | 91 | 8 | 1 | 0 | 91.9% (84.9–95.8) | <0.001 |
| | CT−/GPT+ (n=50) | 47 | 2 | 1 | 0 | 95.9% (86.3–98.9) | <0.001 |
| | CT+/GPT− (n=50) | 44 | 6 | 0 | 0 | 88.0% (76.2–94.4) | <0.001 |

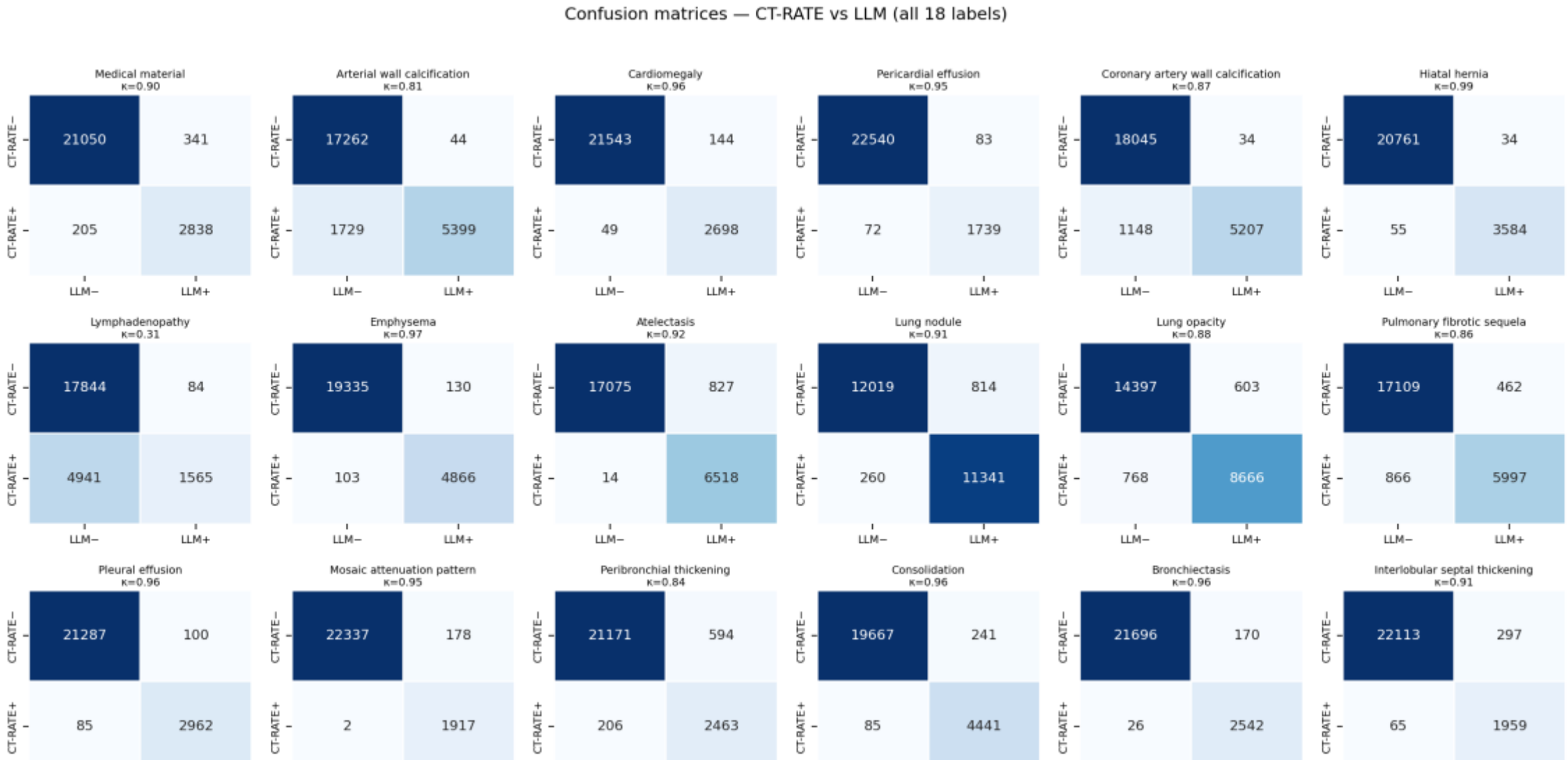

**Supplementary Figure 1.** Per-label confusion matrices comparing CT-RATE labels and GPT-5.4-derived labels across the 18 abnormality categories. For each category, the four cells show the number of reports classified as CT-RATE negative/LLM negative, CT-RATE negative/LLM positive, CT-RATE positive/LLM negative, and CT-RATE positive/LLM positive. Cohen's κ for each category is shown above each panel.